\documentclass{article}
\usepackage[utf8]{inputenc}
\usepackage{amsmath,amssymb}
\usepackage{graphicx}
\usepackage[a4paper,top=3cm,bottom=2cm,left=3cm,right=3cm,marginparwidth=1.75cm]{geometry}

\title{Cosmic ray anisotropies from transient extragalactic sources}
\author{Diego Harari, Silvia Mollerach and Esteban Roulet\\
Centro At\'omico Bariloche, Comisi\'on Nacional de Energ\'\i a At\'omica\\ 
Consejo Nacional de Investigaciones Cient\'\i ficas y T\'ecnicas (CONICET)\\ 
Av. Bustillo 9500, R8402AGP, Bariloche, Argentina}
\date{}

\begin{document}
\maketitle
\begin{abstract}
We study the spectrum and anisotropies of ultrahigh energy cosmic ray transient sources, accounting for the effects of their propagation through the turbulent extragalactic magnetic fields. We consider either bursting sources or sources emitting since a given initial time. We analyse in detail the transition between the diffusive and the quasi-rectilinear regimes, describing some  new features that could be present. 
\end{abstract}

\section{Introduction}

The sources of the ultrahigh energy cosmic rays (UHECRs)  are still unknown, but the expectation is that one may eventually be able to identify them  through the study of the anisotropies in the distribution of their arrival directions. The main difficulty that appears is that, cosmic rays (CRs) being charged nuclei, their trajectories get deflected by the Galactic and extragalactic magnetic fields that they traverse as they travel to us, and hence their arrival directions do not point towards their sources. However, the deflections decrease for increasing rigidities (which is the momentum per unit charge), and they may become smaller than a few tens of degrees  at the highest observed energies. This gives the hope that one may be able to infer the location of the closest powerful extragalactic sources by identifying excesses in the CR arrival directions around them. Besides the distribution in the sky of the  arrival directions, also the energy dependence of the observed patterns and the detailed evolution with energy of the CR mass composition are important for this search. The eventual separation of light and heavy components, which suffer different amounts of deflection, could also be helpful in this respect, and this is something that will be exploited by the ongoing upgrade of the Pierre Auger Observatory.

Another ingredient that is relevant in the search for the CR origin is the fact that one does not expect that the sources be steady. Although the steadiness of the sources is the simplest assumption that is usually considered, all candidate sources have some degree of variability. In particular, among the plausible UHECR candidate sources are gamma ray bursts (GRBs), which have a prompt emission  taking place on timescales of seconds and an afterglow on timescales of hours to weeks; tidal disruption events (TDEs) are transient events in which the CR acceleration could take place on time scales of weeks to months; active galactic nuclei (AGNs), which  may last for more than $10^7$~yr but their activity  gets enhanced in episodes  of increased accretion or during galaxy mergers, that also promote star formation activity, and variability in their electromagnetic flux on timescales of days to years has been observed. 

Scenarios with one or a few transient sources dominating the CR spectrum at the highest energies \cite{mr19} also provide an attractive option to account for the apparently hard source spectrum that is inferred from spectral and composition observations \cite{combfit}, where the observed hardness may be associated with a propagation effect rather than a characteristic of the source spectral shape.

The main purpose of this work is to study in detail  the  implications  of the variability of the CR sources  on the potentially observable anisotropies, focusing on the high energy regime in which there is  a transition between the diffusive and ballistic CR propagation in the turbulent extragalactic magnetic fields. 
 We extend to the case of transient sources the characterization of the distribution of arrival directions in different propagation regimes that was performed in ref.~\cite{hmr16} for the case in which the sources are emitting steadily since infinite time in the past.

\section{CR propagation in turbulent magnetic fields}

We aim to describe the arrival directions of UHECRs from extragalactic sources in our cosmic neighbourhood, within at most about one hundred Mpc, which could be the origin of localized CR excesses in the sky. We will thus neglect attenuation effects upon the energy and composition of UHECRs, such as photo-pion production by protons, pair production losses or  photo-disintegration of nuclei during their propagation. We will analyze the impact of their propagation across turbulent extragalactic magnetic fields as they travel from their sources towards Earth. The CR propagation will thus be mostly affected by the magnetic fields within the Local Supercluster region, which are larger  than the average value over the whole universe. In particular, large-scale inhomogeneities such as those associated with the cosmic voids will be ignored in this study.

We will hence consider the idealized situation of CR propagation in a homogeneous and isotropic  turbulent extragalactic magnetic field. In this case, there is a critical energy that separates different regimes of CR propagation:
\begin{equation}
E_{\rm c}= ZeBl_{\rm c}\simeq 0.9Z\frac{B}{\rm nG}\frac{l_{\rm c}}{\rm Mpc}\,\text{EeV}.
\end{equation}
This is the energy for which the effective Larmor radius $r_L=eZE/B$ coincides with the coherence length $l_{\rm c}$ of the magnetic turbulence having root mean square (rms) strength $B$. If $E<E_{\rm c}$, the deflections imprinted by the magnetic field modes with wavelength comparable to the Larmor radius are large and there is resonant diffusion. If $E>E_{\rm c}$, the deflections across each coherent domain are small, and the total deflection becomes sizable only after the CRs traverse several of them. The distance scale over which the deflection becomes of  order $\sim 1$~rad is known as the diffusion length $l_D$. At distances sufficiently larger than $l_D$ the propagation enters the regime of spatial diffusion, characterized by an isotropic diffusion coefficient $D$ such that $l_D\equiv 3D/c$. If the source distance is comparable or smaller than $l_D$ the propagation is instead quasi-rectilinear.

The energy dependence of the diffusion length is a crucial ingredient to analyze the propagation of CRs across turbulent magnetic fields. We have evaluated in \cite{difu2} the energy dependence of the diffusion coefficient $D(E)$ through numerical integration of the trajectories of charged particles in a homogeneous turbulent magnetic field.  In the present work we will model for definiteness the turbulence of the extragalactic magnetic field with a Kolmogorov spectrum, such that the field energy density scales as $\omega(k)\propto k^{-5/3}$ in Fourier space. The analytic fit to $D(E)$ obtained in \cite{difu2} (see also \cite{gl07,difu1}) is given in such a case by
\begin{equation}
D(E) \simeq \frac{c}{3}l_{\rm c}\left[ 4\left(\frac{E}{E_{\rm c}}\right)^2 + 0.9\left(\frac{E}{E_{\rm c}}\right) + 0.23\left(\frac{E}{E_{\rm c}}\right)^{1/3}\right] \ .
\label{D(E)}
\end{equation}
The energy dependence can also be evaluated for other types of turbulence, such as for instance one with a Kraichnan distribution.

Our analyses will be performed in terms of $E/E_{\rm c}$ for given source distances and duration of their emissivity. Precise values of the extragalactic magnetic field parameters are not known, and they  likely depend upon source location. Realistic estimates range around 1--100~nG for their rms strength in the Local Supercluster region, and the coherence length may range from 10~kpc to 1~Mpc (see e.g. \cite{vallee11,fe12,enzo17}).  

Our aim is to characterize the spectrum and angular distribution of the CRs that reach Earth from a transient source at distance $r_{\rm s}$ after propagation in a turbulent magnetic field. We do so following the method implemented in \cite{hmr16}: a numerical integration of the stochastic differential equation that describes the scattering of UHECRs in a turbulent homogeneous and isotropic magnetic field  \cite{ac99}
\begin{equation}
{\rm d}n_i =-\frac{1}{l_D}n_i c{\rm d}t + \frac{1}{\sqrt{l_D}} P_{ij}{\rm  d}W_j,
\label{dni}
\end{equation}
where $P_{ij} \equiv (\delta_{ij} - n_i n_j)$ is the projection tensor onto the plane orthogonal to the direction of the CR velocity given by $\hat n\equiv (n_1,n_2,n_3)$,
repeated indices are summed and (${\rm d}W_1,\,{\rm d} W_2,\, {\rm d}W_3$) are three Wiener processes such that $\langle{\rm d} W_i \rangle =0$ and  $\langle {\rm d}W_i{\rm d}\,W_j \rangle =c\,{\rm d}t\,\delta_{ij}$.
Implementing this method, we have characterized in \cite{hmr16} the distribution of arrival directions in different propagation regimes for UHECRs originated from steady sources active since infinite time in the past. In the present work we will implement the same formalism to analyze the angular distribution from the diffusive to the ballistic regimes of UHECRs emitted by transient sources, both of a bursting nature as well as those steadily emitting since a given finite initial time in the past.

Note that the approach described above does not consider a fixed realisation for the turbulent magnetic field, but rather averages over possible deflections in random realisations, and hence it reproduces the general expected features of the diffusion process. However, the specific details of the deflections may differ in a given realisation when the propagation is almost rectilinear, in particular when the maximum transverse deflection between alternative trajectories becomes comparable or smaller than the coherence length $l_{\rm c}$.  In this latter case, a different approach would be required, and one expects to observe separate multiple images of the source, with potentially strong energy dependent magnifications of their fluxes, as was discussed in \cite{lensing}. 
We also note that besides the extragalactic turbulent fields, CRs have to traverse  the Galactic magnetic field, which has both a turbulent and a regular component. The effects of the turbulent Galactic field is however expected to be smaller than those of the extragalactic fields considered here, given that the former has a much smaller spatial extent.\footnote{In order that the deflection in the extragalactic turbulent magnetic field that we consider  be larger than the deflection in the turbulent Galactic magnetic field, of strength $B_{\rm g}$ and coherence length $l_{\rm c}^{\rm g}$, one needs that $B/{\rm nG}>0.1(B_{\rm g}/\mu{\rm G})\sqrt{(10\,{\rm Mpc}/r_{\rm s})(100\,{\rm kpc}/l_{\rm c})(L/{\rm kpc})(l_{\rm c}^{\rm g}/10\,{\rm pc})}$, with $L$ the distance traversed through the Galactic turbulent field and $r_{\rm s}$ the distance to the extragalactic source.} 
At the high energies considered here, the regular Galactic field will contribute mainly to a global energy dependent coherent deflection of the images, in an amount and direction depending on the arrival direction considered, which can in principle be accounted for separately (their size is typically of order one degree for protons with an energy of 100~EeV). Large scale coherent extragalactic magnetic fields,  in case they were to exist, could further contribute to these deflections.

\section{The case of a bursting source}

Let us start by considering a source at a distance $r_{\rm s}$ that emits CRs during a brief period of time, with duration negligible with respect to the time for straight propagation from the source to the observer, so that one may consider the emission to be a burst (see  \cite{be90,mi96,wa96} for some initial studies on this subject). If the emission happened a time $t$ before the observation, so that the distance traveled by the CRs along their trajectory is $ct$, we will denote $d\equiv ct/l_D$, which is the distance traveled in units of the diffusion length (note that the latter  is energy dependent). We will similarly consider the distance from the source in units of the diffusion length as $R\equiv r/l_D$, and the  predictions can then be conveniently expressed in terms of $d$ and $R_{\rm s}=r_{\rm s}/l_D$. For the CRs to be able to reach the Earth one clearly needs that $d>R_{\rm s}$ (i.e. $ct>r_{\rm s}$).

In the spatial diffusive regime that applies when the distance travelled is much larger than the diffusion length ($d\gg 1$), the CR density as a function of the distance from the source  is generally described by the solution of the diffusion equation
\begin{equation}
    N_{\rm diff}(r,t)=  \frac{N_0}{(4\pi l_Dct/3)^{3/2}}\exp\left[-\frac{3r^2}{4l_Dct}\right],
     \label{nd1}
     \end{equation}
     with the normalization being such that the density  $N_{\rm diff}$ integrates over the whole space to the total number of particles $N_0$ emitted in the burst (for a given differential energy bin). It is convenient to introduce a rescaled density depending just on $R$ and $d$, whose integral over $R$ is unity, through
\begin{equation}
    n_{\rm diff}(R,d)\equiv N_{\rm diff}(r,t)
  \frac{l_D^3}{N_0}= \frac{1}{(4\pi d/3)^{3/2}}\exp\left[-\frac{3R^2}{4d}\right].
     \label{nd1b}
     \end{equation}
     
The above expressions have the drawback that they do not vanish for $r>ct$ (i.e. for $R>d$), implying an unphysical `superluminal' motion. A possible fix to this problem was proposed in ref.~\cite{du07}, relying on the relativistic J\"utner propagator \cite{juttner} (and generalized to the case including energy losses in \cite{aloisio}), through the expression
\begin{equation}
     N'_{\rm diff}(r,t)=\frac{3N_0}{8\pi (ct)^2K_1(1.5ct/l_D)l_D}\exp\left[-\frac{3ct}{2\sqrt{1-(r/ct)^2}}\right]\frac{1}{[1-(r/ct)^2]^2},
     \label{nd2}
\end{equation}
 with $K_1$ being the modified Bessel function. Note that in the limit $ct\gg r$ one has that $N'_{\rm diff}\to N_{\rm diff}$.
It is useful to also introduce in this case a rescaled density through
\begin{equation}
     n'_{\rm diff}(R,d)=N'_{\rm diff}(r,t)
  \frac{l_D^3}{N_0}=\frac{3}{8\pi d^2K_1(1.5d)}\exp\left[-\frac{3d}{2\sqrt{1-(R/d)^2}}\right]\frac{1}{[1-(R/d)^2]^2}.
     \label{nd2b}
\end{equation}

\begin{figure}[t]
\centering
\includegraphics[scale=1,angle=0]{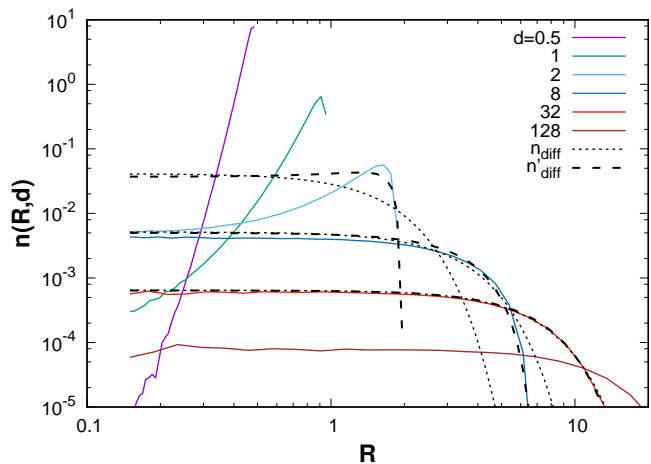}
\caption{Density distribution as a function of R for different values of $d$. For the illustrative cases $d=2$, 8 and 32 we also show the curves that correspond to the distributions in eqs.~(\ref{nd1b}) and (\ref{nd2b}). }
\label{f1}
\end{figure}

In Fig.~\ref{f1} we show the CR densities,  obtained through simulations of a large number of CR trajectories computed by solving the stochastic equations, for several values of $d$. For the illustrative cases $d=2$, 8 and 32 we also displayed the curves corresponding to the expressions in eqs.~(\ref{nd1b}) and (\ref{nd2b}). For the cases shown with $d\geq 8$ the good agreement obtained with eq.~(\ref{nd2b}) is apparent,  while eq.~(\ref{nd1b}) is only accurate  when the distance from the source is much smaller than the distance travelled, i.e. for $R\ll d$. 
However, for smaller values of $d$ the match with the analytic expression is not good, and the disagreement becomes  more pronounced as the value of $d$ decreases.
The main reason for this is that for $d<2\pi$ the typical CR trajectories do not manage to make more than one whole turn (remember that $l_D$ is the distance over which the particle deflections are of order of 1~rad), and hence in this case there is still a strong memory of the initial velocity direction that the particles had when they exited the source. This translates into a density distribution with the shape of an inflating balloon that gets progressively thicker and eventually dissolves into the flatter profile associated with the diffusive regime, as is seen in Fig.~\ref{f1}. We will discuss here in detail the main features of this initial period, that we refer to as the prompt phase. The study of this phase is particularly relevant because it is in this regime that one expects that the CR flux excesses could become more localized in the sky.\footnote{See  \cite{fedorov} for a different approach to include a ballistic regime in terms of a sub-population of particles which are assumed not to be scattered by the magnetic fields. }

\begin{figure}[t]
\centering
\includegraphics[scale=1,angle=0]{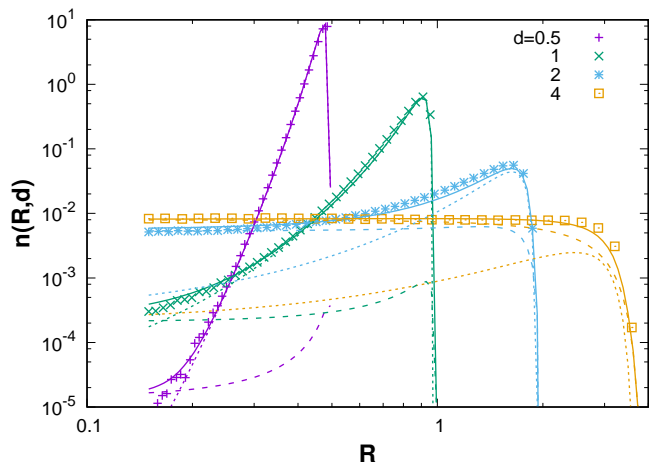}
\caption{Density distribution as a function of $R$ for different values of $d\leq 4$. Also shown are the profiles expected according to eq.~(\ref{nall}) and with dotted lines the individual contribution from the prompt component, while with dashed lines that of the diffusive component.}
\label{fprompt}
\end{figure}

Although we do not have an exact solution for this transient phase, we provide here an analytic fit to the results of the simulations that is quite accurate and is helpful to understand the main features of this regime. We model the prompt contribution to the density as a Gaussian centered at a radius $\bar R(d)$ and with dispersion $\sigma_R(d)$, conveniently distorted so as to ensure the absence of superluminal particles (in the spirit of \cite{juttner}). In this phenomenological approach, the prompt CR density is taken as
\begin{equation}
    n_{\rm prompt}(R,d)=n_1(d)\exp\left[-\frac{(R-\bar R)^2}{2\sigma_R^2(1-(R/d)^2)^\alpha}\right]\frac{1}{[1-(R/d)^2]^{1.5}},
     \label{np}
\end{equation}
with $\alpha=1.25-0.1/d$.
To account for the $d$ dependence of $\bar R$, we exploit the analytic solution for $\langle R^2\rangle$ derived in \cite{ac99}, where it was found that
\begin{equation}
    \langle R^2\rangle=2[d-1+\exp(-d)] ,
\end{equation}
and adopt $\bar R=\sqrt{\langle R^2\rangle}$ (no analytic solution for $\langle R\rangle$ is available). For  $\sigma_R$ we exploit the knowledge of the dispersion of the particles along the direction of their initial velocities, denoted as $\sigma_z= \sqrt{\langle z^2\rangle-\langle z\rangle^2}$, with \cite{ac99}
\begin{equation}
    \langle z^2\rangle=\frac{2}{3}\left(d-\frac{1}{3}[1-\exp(-3d)]\right)
\,\,\, , \,\,\,\langle z\rangle=1-\exp(-d) .
\end{equation}
The actual dispersion in the radial direction is expected to be qualitatively similar but smaller than that along the initial velocity direction, 
and we have found that a good fit to the results is obtained just setting $\sigma^2_R=0.75 \sigma^2_z$. The normalization factor $n_1(d)$ is obtained requiring that
\begin{equation}
    4\pi \int_0^{d} n_{\rm prompt}(R,d)\, R^2{\rm d}R=1.
\end{equation}

Finally, the total rescaled density will be the weighted sum of the prompt and diffusive contributions,
\begin{equation}
  n(R,d) =f(d)n_{\rm prompt}(R,d)+[1-f(d)]n_{\rm diff}(R,d) ,
  \label{nall}
\end{equation}
where $f(d)$ is the fraction of the emitted particles that are described by the prompt density profile at the time  parametrized by $d$. One expects that $f\to 1$ for $d\ll 1$ while $f\to 0$ for $d\gg 2\pi$.
The fitted density profiles for the values $d=0.5$, 1, 2 and 4 are shown in Fig.~\ref{fprompt}, and the agreement with the simulations is quite good in all cases.

\begin{figure}[t]
\centering
\includegraphics[scale=1,angle=0]{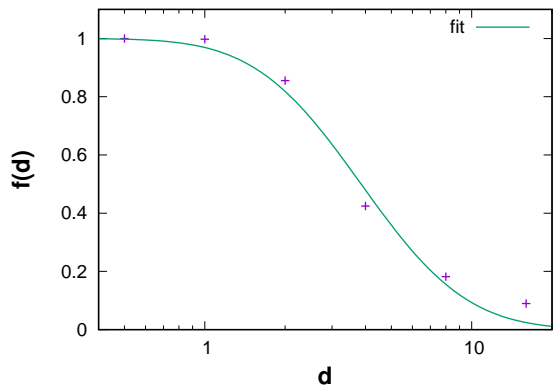}
\caption{Fraction of the prompt contribution $f(d)$ as a function of $d$, together with the fit (see text). }
\label{ffd}
\end{figure}

 A plot of the values of the fraction $f$ associated to the prompt component, together with an analytic fit of the form $f(d)=[1+{\rm Erf}(0.59-\log(d))/0.44]/2$, is shown in Fig.~\ref{ffd}.

\begin{figure}[h]
\centering
\includegraphics[scale=.97,angle=0]{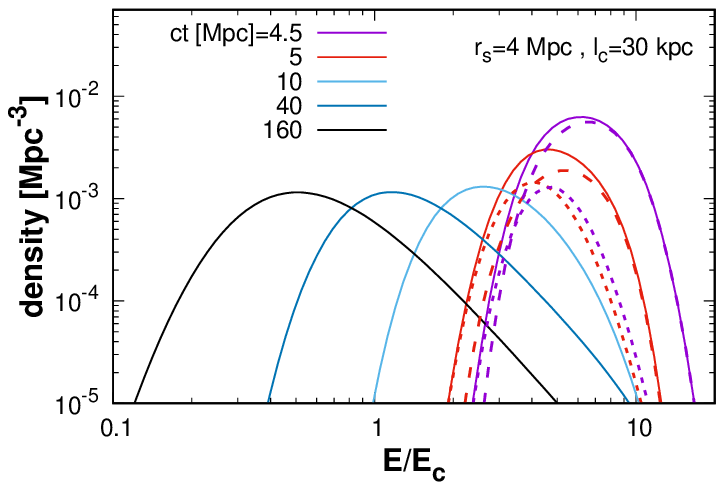}
\includegraphics[scale=.97,angle=0]{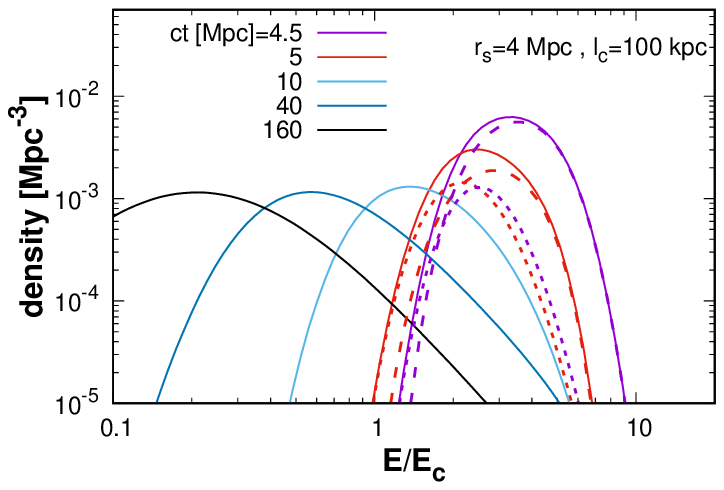}
\caption{Normalized CR density  as a function of $E/E_{\rm c}$ for a source at $r_{\rm s}=4$~Mpc. For the smaller values of $ct=4.5$ and 5~Mpc, the long dashed lines indicate the contribution from the prompt component, and short dashes that from the diffusive component. For the larger values of $ct$ shown the prompt component becomes negligible. Left panel is for $l_{\rm c}=30$~kpc, right panel for $l_{\rm c}=100$~kpc.}
\label{fne}
\end{figure}

To apply the previous results to a specific physical situation, we consider a scenario in which one has a source at a distance of 4~Mpc, which is similar to the distance to the nearby AGN Centaurus~A. Given the uncertain coherence length and strength of the turbulent extragalactic magnetic field, we consider two values $l_{\rm c}=30$~kpc and 100~kpc, and give the results in terms of the energy ratio $E/E_{\rm c}$ for different values of $ct$. Figure~\ref{fne} shows the CR density that would be observed at the Earth, normalized to the injection of one particle in a given energy bin (equivalently this would be the density for a flat energy spectrum). For the smaller values of $ct$ we also show separately the prompt and diffusive contributions. Note that the fraction $f(d)$ is non-negligible only for $d<10$ (Fig.~\ref{ffd}), which approximately corresponds to $E/E_{\rm c}>3\sqrt{ct/10\,{\rm Mpc}}\sqrt{30\,{\rm kpc}/l_{\rm c}}$. One should keep in mind that $f_d$ is the fraction of the prompt component after integration over all space, but the fraction contributed by the prompt flux is not uniform in space and depends on the actual distance from the observer to the source, $r_{\rm s}$. In particular, the prompt fraction at the observer location  may become actually much smaller than $f(d)$ if $ct-r_{\rm s}\gg l_D$. 
One can see that the main effect of increasing the coherence length for a fixed source distance is to shift the spectrum to smaller values of $E/E_{\rm c}$, with the shift scaling approximately as $l_{\rm c}^{-1/2}$ (since for $E>E_{\rm c}$ one has that $l_D\propto l_{\rm c}(E/E_{\rm c})^2$), but for increasing $l_c$ the distributions become also wider in logarithmic scale for   $E<E_{\rm c}$,  in which case one approaches the resonant diffusion regime for which $l_D\propto l_{\rm c}(E/E_{\rm c})^{1/3}$.

The actual differential density $n(E)$ can be found by multiplying the normalized density discussed above by d$N/{\rm d}E$, that is the number of CRs emitted by the burst in a given energy interval. 
In particular, for a power-law spectrum such that d$N/{\rm d}E\propto E^{-\alpha}$, one would have that $E^\alpha n(E)$ will have a similar shape as the normalized densities shown in Fig.~\ref{fne}. Note that once the turbulent magnetic field parameters $B$ and $l_{\rm c}$ are fixed, the variable $E/E_{\rm c}$ is just proportional to the rigidity $E/Z$ of the particles. Thus, if the source is emitting a mixed composition of nuclei with various charges $Z$, the heavier nuclei will have similar spectra as protons, shifted to the right by a factor $Z$. 
In the region where particles are diffusing, the maximum of the distribution is reached for an energy $E_{\rm max}$ such that $l_D(E_{\rm max}/E_{\rm c})\simeq r_{\rm s}^2/2ct$ \cite{mr19}. 
In particular, if the maximum appears at energies larger than $E_{\rm c}$, so that $l_D\simeq 4l_{\rm c}(E/E_{\rm c})^2$, one would get $E_{\rm max}\simeq E_{\rm c}\sqrt{r_s^2/8l_{\rm c}ct}$.
When the prompt phase gives a relevant contribution, the maximum is slightly shifted to larger energies, as  can be seen in Fig.~\ref{fne}.

\begin{figure}[h]
\centering
\includegraphics[scale=0.8,angle=0]{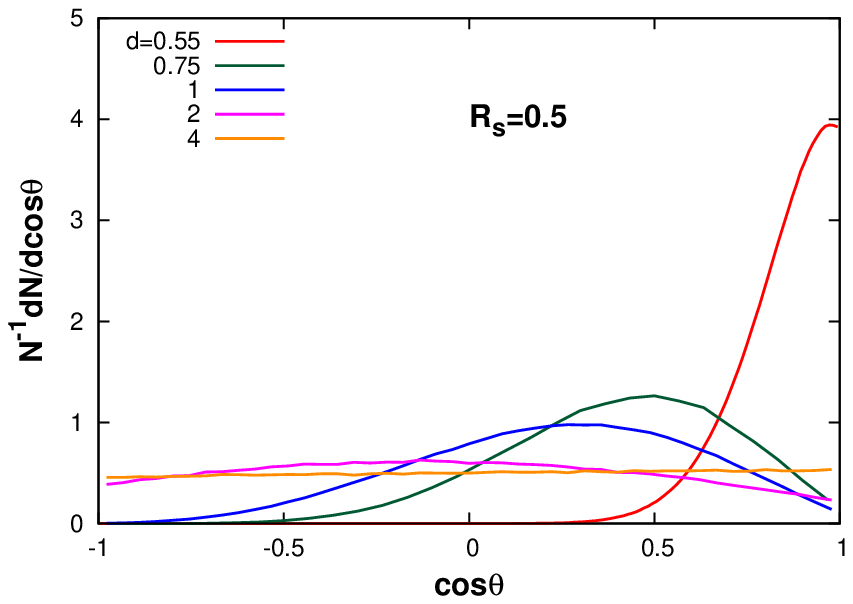}
\includegraphics[scale=0.8,angle=0]{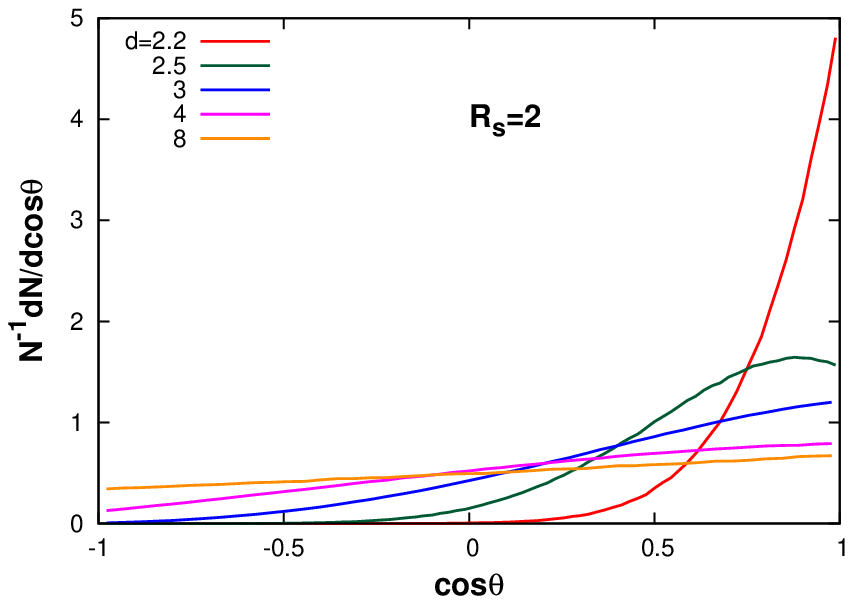}

\caption{Distribution of CR arrival directions as a function of $\cos\theta$ for values of the normalized source distance $R_{\rm s}\equiv r_{\rm s}/l_D=0.5$ and 2. The lines correspond to different values of $d$, which is the distance traveled by the particles since the emission of the burst, in units of $l_D$. When $d/R$ is larger than a few, the distribution approaches that of a dipole with amplitude $R/2d$.}
\label{fig:burst.dist}
\end{figure}

\begin{figure}[h]
\centering
\includegraphics[scale=0.8,angle=0]{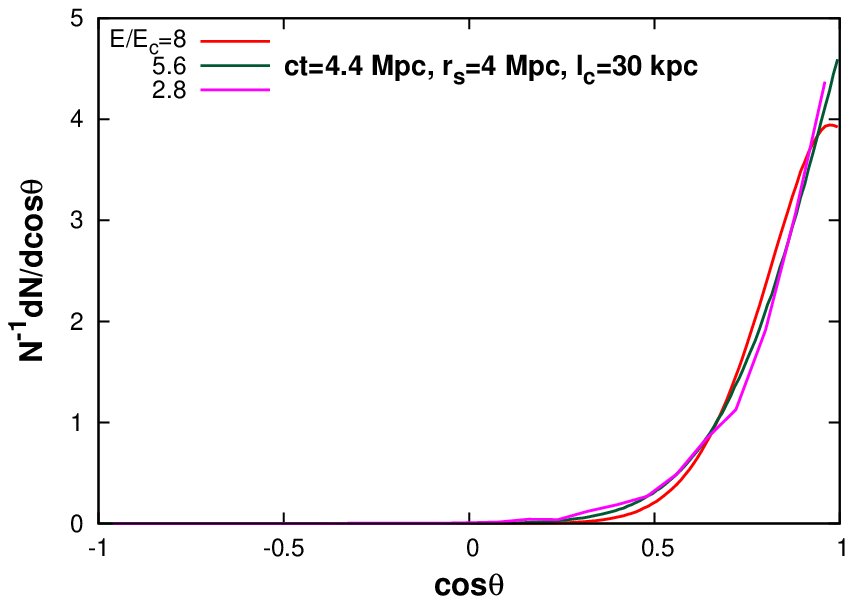}
\includegraphics[scale=0.8,angle=0]{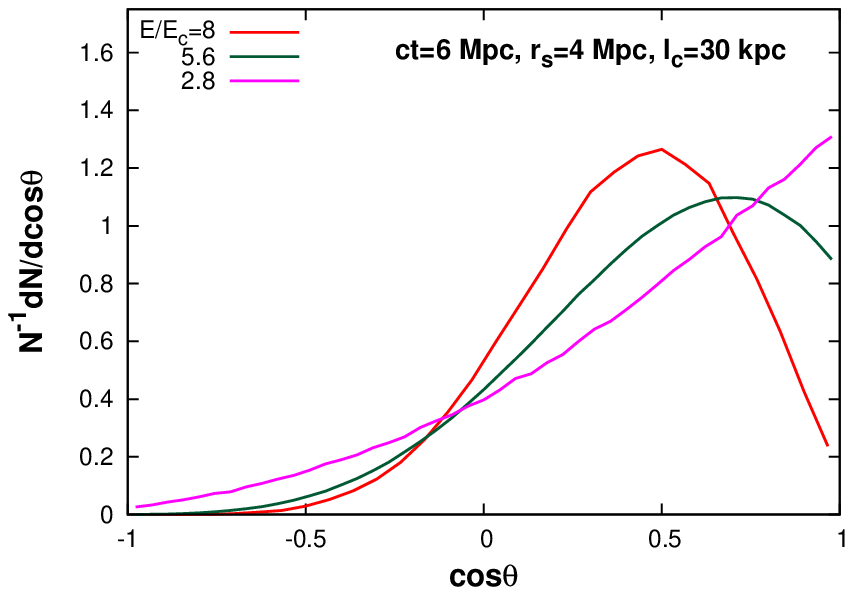}
\includegraphics[scale=0.8,angle=0]{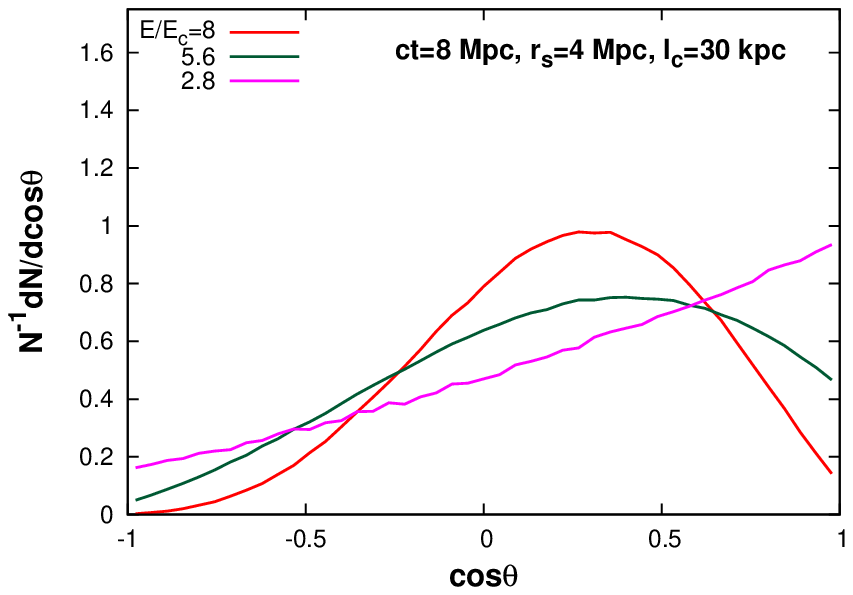}
\includegraphics[scale=0.8,angle=0]{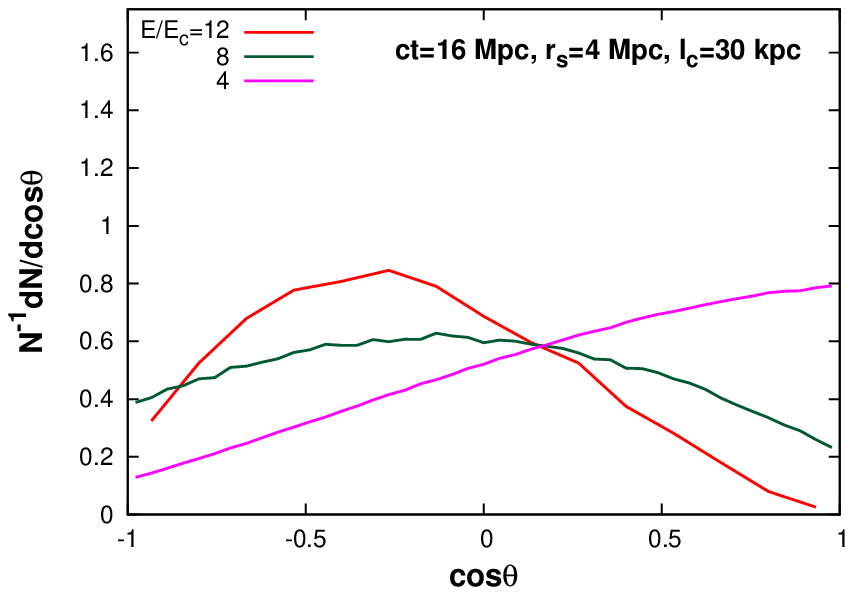}
\caption{Distribution of CR arrival directions as a function of $\cos\theta$ for a source at a distance $r_{\rm s}=4$~Mpc that had a burst at a time such that $ct=4.4,6,8$ and $16$~Mpc, as labeled. The lines correspond to different values of $E/E_{\rm c}$ within an interesting range in which the propagation is transitioning away from rectilinear.}
\label{fig:burst.dist4Mpc}
\end{figure}

The other important feature of the CR flux is the distribution of arrival directions at a given source distance. This is depicted in Fig.~\ref{fig:burst.dist}, where the distribution in $\cos\theta$ is shown, with $\theta$ being the angle with respect to the source direction. This distribution is displayed for two different source distances $R_{\rm s}=0.5$ and 2, and for different values of $d$. 
By comparing the different curves in each plot, one can see that for $d/R_{\rm s}<1.3$ the distribution is very peaked in the  forward direction, since in this case one would be directly observing the passage of the front of the prompt shell of CRs. However, one also finds that for $R_{\rm s}<2$, in which case the propagation is quasi-rectilinear,  the distribution starts to become suppressed in the direction towards the source for $d-R_{\rm s}\simeq 0.2$ to 1, because in this case the particles having the straighter trajectories would have already passed through the Earth in the far past, while those that do reach the observer are subject to increasingly large average deflections as $d$ increases, so that they arrive preferentially sideways. When $R_s$ is small, as can be seen in the case with $R_{\rm s}=0.5$, the flux can be suppressed in the direction to the source for $d\leq 2$, and it actually may get enhanced in the  hemisphere opposite from that of the source, as is apparent in the case shown with $d=2$. In the regime with $d-R_{\rm s}\gg 1$  the distributions  flatten due to the contribution of particles that made more than one turn along their trip. The main feature in this regime is the presence of a dipolar component in the flux distribution, whose amplitude is given by $\Delta\simeq 1.5~ R_{\rm s}/d=1.5~r_{\rm s}/ct$, which is actually independent of the energy considered \cite{be90,fa00}. 

In Fig.~\ref{fig:burst.dist4Mpc} we illustrate the features described above within a specific physical scenario, assuming a source 4~Mpc away and an extragalactic magnetic field with coherence length $l_{\rm c}=30$~kpc. 
The distributions of arrival directions are shown for bursts for which the CR travel time was factors 1.1, 1.5, 2 and
4 times larger than the straight trajectory travel time.
 They are displayed for different values of the particles rigidities, in terms of $E/E_{\rm c}$. We note that for the chosen source distance and coherence length the energy at which the rms deflection is 1~radian corresponds to $8.5E_{\rm c}$. The top-left panel illustrates that if the burst occurred at a time only slightly larger than the time needed for rectilinear propagation, then the distribution does not change appreciably with energy. This is so because only almost rectilinear trajectories can reach Earth within such relatively short time.  While the distribution does not change appreciably, clearly the fraction of trajectories that can arrive does significantly decrease with energy, as was shown in Fig.~\ref{fne}. The subsequent panels, considering bursts that occurred at increasingly larger times in the past, illustrate the energy-dependence of the distributions and their different features. Since the CRs with quasi-rectilinear trajectories have already passed by, the distributions at the highest energies shown are peaked at increasingly sideways directions and then to backwards arrivals for earlier bursts. For lower energies the distributions flatten due to the spatial diffusion.  At comparable energies the flattening is more pronounced for earlier bursts, since in this case there was more time available for the particles to diffuse.

\begin{figure}[h]
\centering
\includegraphics[scale=0.85,angle=0]{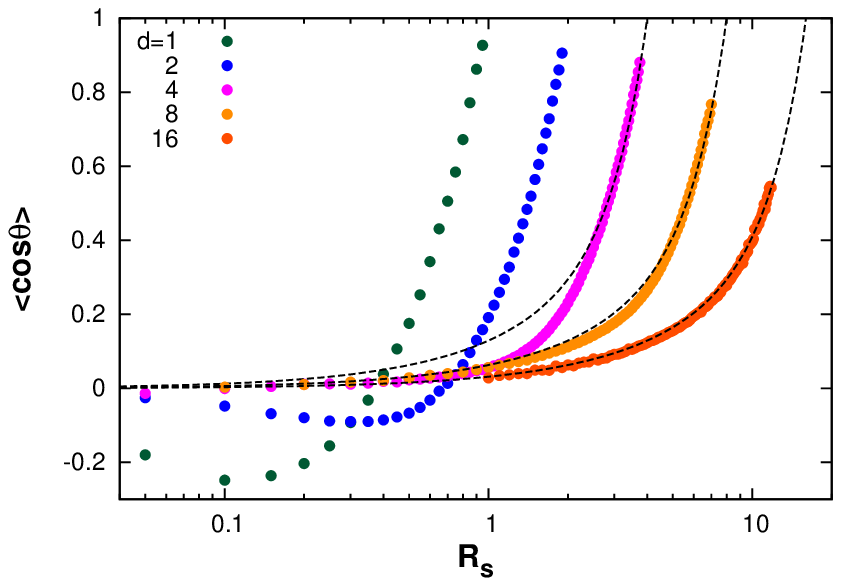}\includegraphics[scale=0.85,angle=0]{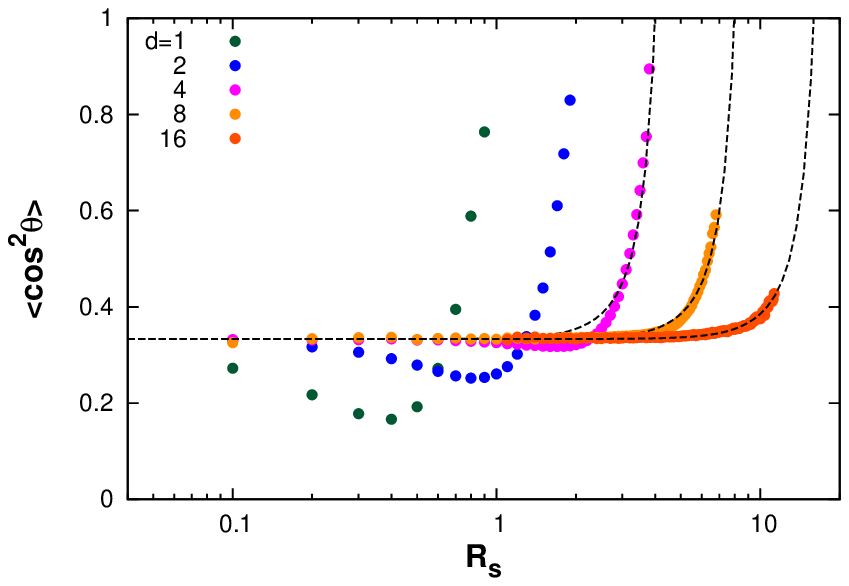}
\caption{Values of $\langle\cos\theta\rangle$ (left) and $\langle\cos^2\theta\rangle$ (right), as a function of the normalized source distance $R_{\rm s}$, for different values of the time of emission of the burst as indicated. The black dashed lines in the left plot correspond to the function $(R_{\rm s}/2d)(1+(R_{\rm s}/d)^{2.5})$, which is a good fit for $d > 8$ for all values of $R_{\rm s}$. For $d < 8$ it is a good fit for $R_{\rm s} >d/2$ only. In the right plot the black dashed lines  correspond to the fit by the function $
(1+(R_{\rm s}/d)^4(1+(R_{\rm s}/d)^8))/3$.}
\label{fig:burst.cos.16}
\end{figure}

The average values of $\cos\theta$ as a function of $R_{\rm s}$, for different values of $d$, are shown in Fig.~\ref{fig:burst.cos.16} (left panel). The features described above are apparent also in these plots, and one can see that for values of $d>8$ a very good fit to the results is obtained with $\langle\cos\theta\rangle\simeq R_{\rm s}/(2d)[1+(R_{\rm s}/d)^{2.5}]$ in all the range $R_{\rm s}<d$. In the right panel of  Fig.~\ref{fig:burst.cos.16} we show the  average values of $\cos^2\theta$ as a function of $R_{\rm s}$, for different values of $d$, together with the fits with the function $\langle\cos^2\theta\rangle= (1+(R_{\rm s}/d)^4(1+(R_{\rm s}/d)^8))/3$, which accurately reproduces the results of simulations for $d>8$. 

Note that in the diffusive regime, the dipolar component of the distribution is characterised by $\Delta\simeq 3\langle\cos\theta\rangle$, while the quadrupolar term by $q\simeq 45/4(\langle\cos^2\theta\rangle-1/3)$, hence their ratio is approximately given, in the limit $d\gg R_{\rm s}$, by $q/\Delta\simeq 7.5(R_{\rm s}/d)^3$.

\section{The case of a source emitting steadily since a given time}

The ideal case of a steady source emitting since an infinite time leads to a distribution of particles which is independent of time. The angular distribution of the observed particles depends only on the ratio of the source distance to the diffusion length, $R_{\rm s}$, as has been described in \cite{hmr16}.
It  was shown there that a good fit to the  angular distribution obtained in numerical simulations of particle trajectories with stochastic deflections is given by a Fisher distribution characterized by a concentration parameter $\kappa$, describing how much the deflections have dispersed the arrival directions from the source position, plus an isotropic contribution characterized by a parameter $i$, measuring the fraction of particles that diffused for very long times and thus arrive almost isotropically distributed.  The angular distribution is given by
 \begin{equation}
 \frac{1}{N}\frac{{\rm d}N}{{\rm d}\cos\theta}=\frac{i}{2}+(1-i)\frac{\kappa\exp(\kappa\cos\theta)}{2\sinh\kappa}
 \label{eq:fisher}
 \end{equation}
     
The first two moments of this distribution are
  \begin{equation}
 \langle\cos\theta\rangle=(1-i)\left(\coth{\kappa}-\frac{1}{\kappa}\right)
 \end{equation}
 and  
 \begin{equation}
 \langle\cos^2\theta\rangle= \frac{i}{3}+(1-i)\left(1+\frac{2}{\kappa^2}-\frac{2}{\kappa \tanh{\kappa}}\right)=1-\frac{2\langle\cos{\theta}\rangle}{\kappa}-\frac{2i}{3}.
 \end{equation}
If the distribution of arrival directions is well characterized by eq.~(\ref{eq:fisher}), any pair of the quantities $\kappa$, $i$, $\langle\cos{\theta}\rangle$ or $\langle\cos^2{\theta}\rangle$ can be used to describe it. The parameters $\kappa$ and $i$ can in fact be obtained from $\langle\cos{\theta}\rangle$ and $\langle\cos^2{\theta}\rangle$ using that
 \begin{equation}
 \frac{2}{3(\coth\kappa-1/\kappa)}-\frac{2}{\kappa}=\frac{\langle\cos^2\theta\rangle-1/3}{\langle\cos\theta\rangle}\equiv\alpha.
 \end{equation}
 An approximate solution to this transcendental equation is given by
 \begin{equation}
 \kappa\simeq \frac{5\alpha-27\alpha^2/4+27\alpha^3/8}{2/3-\alpha}.
 \label{eq:kappa}
 \end{equation}
Finally,
 \begin{equation}
 i=1-\frac{\langle\cos\theta\rangle}{\coth\kappa-1/\kappa}.
 \label{eq:i}
 \end{equation}
 Notice that in a multipolar expansion of the angular distribution (where d$N/$ d$\cos{\theta} \simeq (N/2)(1+\Delta \cos{\theta} + q(\cos^2\theta -1/3)+ \dots)$), the dipolar component satisfies $\Delta=3\langle\cos{\theta}\rangle$ and the quadrupolar component  satisfies $q=(45/4)(\langle\cos^2{\theta}\rangle-1/3)$.
 Thus, $\langle\cos\theta\rangle$ is directly related  to the dipolar component of the anisotropies. On the other hand, $\kappa$ gives a good description of the angular extension of small and intermediate scale anisotropies, with $\langle{\theta^2}\rangle\simeq 2/\kappa$. For large deflections, i.e. for $\kappa\ll 1$, the dipole and quadrupole of the distribution satisfy $\Delta=(1-i)\kappa$ and $q=(1-i)\kappa^2/2$.
   
For a steady source, good fits to both $\langle\cos\theta\rangle$ and $\kappa$ have been obtained in refs. \cite{difu2} and \cite{hmr16} respectively
 \begin{equation}
 \langle\cos\theta\rangle^{\rm steady}(R_{\rm s})= \frac{1}{3R_{\rm s}}\left[1-\exp\left(-3R_{\rm s}-3.5R_{\rm s}^2\right)\right]\equiv C(R_{\rm s}),
 \label{costeady}
 \end{equation}
 \begin{equation}
 \kappa^{\rm steady}(R_{\rm s})\simeq \frac{1}{R_{\rm s}}\left[ 2+\exp\left( -2R_{\rm s}/3-R_{\rm s}^2/2\right)\right].
 \label{eq:kappasteady}
 \end{equation}

 As for a steady source the density of particles reaches a stationary regime in which it does not depend on time, the flux of particles through any sphere around the source has to be the same. Exploiting the spherical symmetry of the problem, we then obtain the general relation
 \begin{equation}
n^{\rm steady}(r,E) 4 \pi c r^2  C(r/l_D) = Q(E)  
 \end{equation}
 with $Q(E)$ the emissivity of the source (differential in energy). For values of $r\ll l_D$, which correspond to small distances from the source and/or very high energies, one has  that $C(r,E) \simeq 1$ and hence the density of particles decreases as $Q(E)/4\pi r^2$ (rectilinear propagation). For larger distances and/or smaller energies,  the diffusion process leads to an enhancement of the density by a factor equal to $1/C(r/l_D)$ with respect to the rectilinear case, and hence there is a direct relation between the density enhancement and the dipolar anisotropies.  
     
 If we consider instead a source that emitted steadily but since a finite time $t_i$ before the observation, so that the maximum distance travelled by the observed CRs is $c t_i$, the density of low energy particles will get suppressed due to the magnetic horizon effect \cite{le04,be06,gl07}, since being their trajectories substantially deflected the low energy particles may have not enough time to reach the observer. The energy $E_{\rm s}$ below which this suppression appears is determined from the relation $l_D(E_{\rm s})\sim r_{\rm s}^2/ct_i$ (i.e. for $d_i\sim R^2$).
 
  \begin{figure}[h]
\centering{
\includegraphics[scale=1,angle=0]{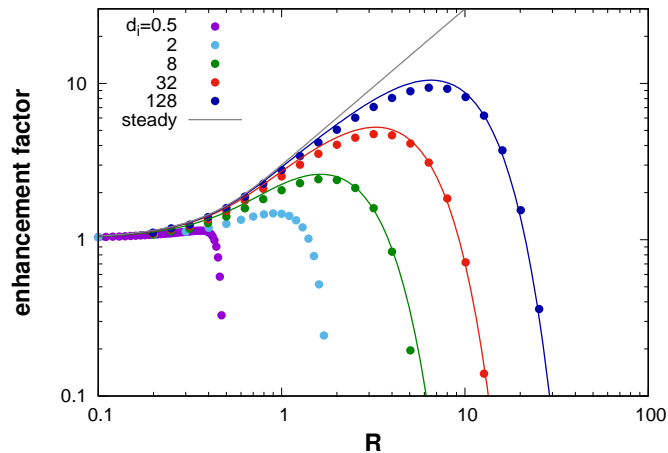}}
\caption{Enhancement factor of cosmic ray density as a function of $R$ for sources emitting since different times, parametrized by the values of $d_i$ quoted. }
\label{fig:nvsr}
\end{figure}
 
The density of particles as a function of the  emission period measured in units of the diffusion length, $d_i\equiv c t_i/l_D(E)$,
 can be written following ref.~\cite{mr19} as 
 \begin{equation}
     n=\frac{Q}{4\pi c r^2}\xi(R,d_i),
 \end{equation}
 where $\xi$ is the enhancement factor in this case. The factor $\xi$  obtained in numerical simulations is shown in Fig.~\ref{fig:nvsr}  as a function  of $R$ and for different values of the emission period $d_i$. For small values of $R$, the factor $\xi$ is close to unity, as expected, while for increasing $R$ the factor  $\xi$ grows as a result of the diffusion enhancement and then drops due to the effect of the magnetic horizon. As the period of emission shortens, less diffusion is possible and thus the enhancement of the density gets smaller. Also the cutoff at large $R$ becomes steeper for decreasing $d_i$, converging to the sharp cutoff of the classical rectilinear propagation horizon at $R=d_i$, as is apparent  for the lowest values of $d_i$  displayed.

 The lines shown in Fig.~\ref{fig:nvsr} correspond to a fit to $\xi$, following ref.~\cite{mr19}, with the expression
 \begin{equation}
     \xi(R,d_i)= \frac{1}{C(R)}\exp\left[-\left(\frac{R^2}{0.6 d_i}\right)^{0.8}\right].
     \label{xi}
 \end{equation}
 For the two shortest periods considered, for which the maximum distance travelled is smaller or equal than twice the diffusion length ($d_i \le 2$),  this diffusion inspired fit is not expected to describe them, and hence the lines are not plotted. In these cases the distribution is actually closer to that expected for rectilinear propagation with a cutoff at $R=d_i$.
 
 \begin{figure}[h]
\centering
\includegraphics[scale=0.8,angle=0]{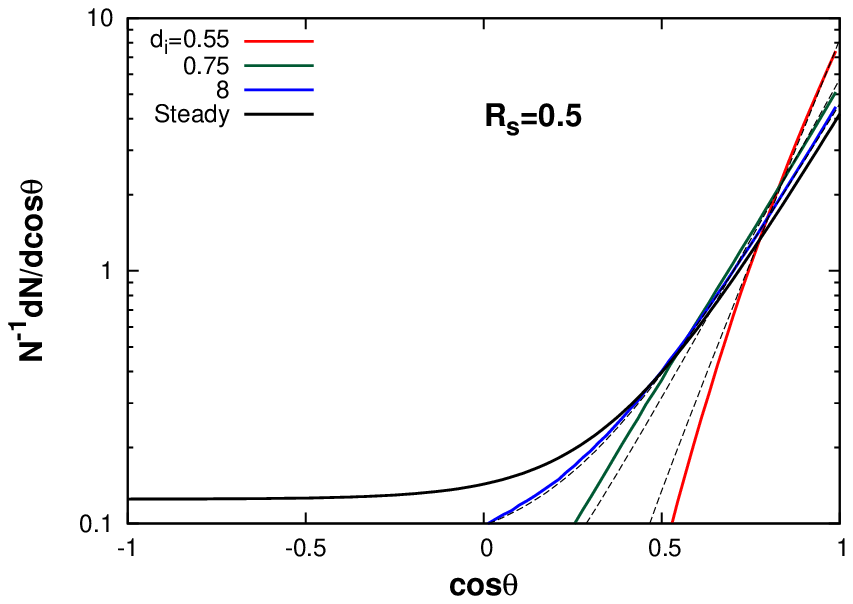}
\includegraphics[scale=0.8,angle=0]{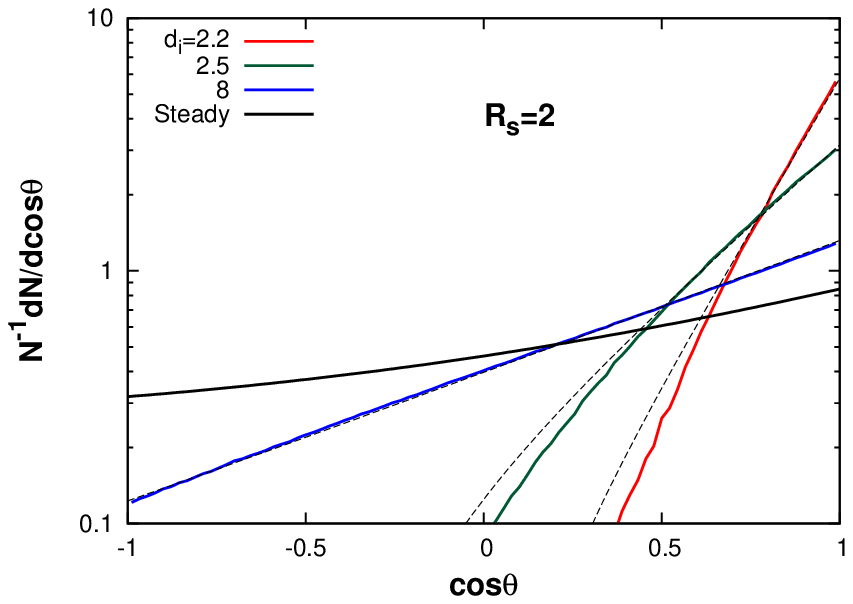}
\caption{Distribution of CR arrival directions as a function of $\cos\theta$ for values of the normalized source distance $R_{\rm s}=0.5$  and 2 for sources emitting since different times, $d_i$ as quoted.  The dashed lines correspond to the distribution in eq.~(\ref{eq:fisher}) with values of  $\kappa$ and $i$ as given by eqs.~(\ref{eq:fitkappa}), (\ref{eq:i}) using (\ref{eq:cosqg2}) or (\ref{eq:cql2}) depending on the value of $d_i$. }
\label{fig:di.dist}
\end{figure}
 
  Regarding the distribution of the arrival directions around the source position, we show it in Fig.~\ref{fig:di.dist} for two different values of the source distance to diffusion length ratio, $R_{\rm s}=0.5$  and 2, and for several values of the duration of the emission period $d_i$. In all cases they are smoothly spread around the source and it turns out that the function in eq.~(\ref{eq:fisher}) provides a reasonably good description, specially in the directions close to the source position. As previously discussed, this can be characterized by the values of $\langle\cos\theta\rangle$ and $\kappa$, and we now analyse the dependence of these quantities on the source distance and duration of the emission period in units of the diffusion length, $R_{\rm s}$ and $d_i$, respectively. 
  
\begin{figure}[h]
\centering{
\includegraphics[scale=0.85,angle=0]{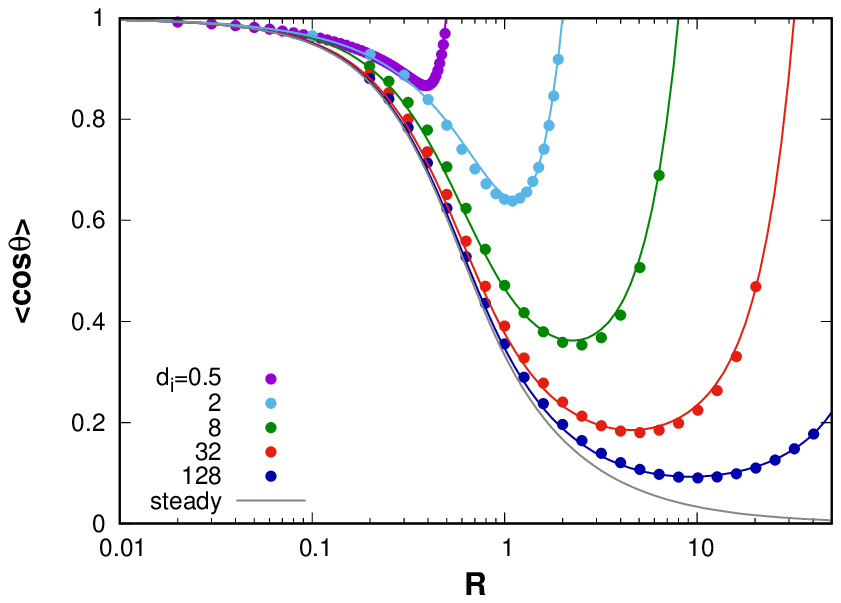}\includegraphics[scale=0.85,angle=0]{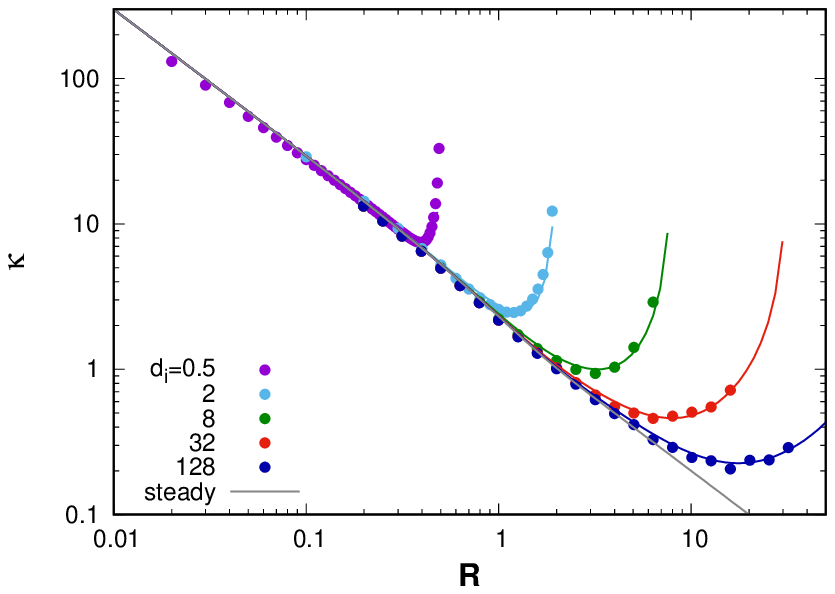} }
\caption{Left: 
Mean value of $\langle\cos\theta\rangle$ obtained through numerical simulations (points) for different values of $d_i$ as labelled in the plot. The solid lines corresponds to the fits from eq.~(\ref{eq:cosqg2}) for $d_i >2$, and from eq.~(\ref{eq:cql2}) for $d_i \le 2$.
Right: Concentration parameter $\kappa$ obtained from simulations using eq.~(\ref{eq:kappa}) (points) and fit from eq.~(\ref{eq:fitkappa}) (lines).}
\label{fig:fitvsr}
\end{figure}

  We show in the left panel of Fig.~\ref{fig:fitvsr} the results for  $\langle\cos\theta\rangle$ as a function of $R_{\rm s}$ for several values of $d_i$. The distribution is more isotropic for longer emission times,  as expected. The arrival directions are very concentrated around the source direction when the diffusion length is much larger than the source distance ($R_{\rm s}\ll 1$) and also for $d_i\rightarrow R_{\rm s}$, in which case only the small fraction of particles emitted at the beginning that suffered the smallest deflections had time to reach the observer. 
  
  From the values of $\langle\cos\theta\rangle$ and $\langle\cos^2 \theta\rangle$ obtained with the simulated particles, the value of $\kappa$
can be obtained from eq.~(\ref{eq:kappa}). This is shown in the right panel of Fig.~\ref{fig:fitvsr} for the same values of $R_{\rm s}$ and $d_i$  reported in the left panel. The concentration parameter is very large for small values of $R_{\rm s}$, as the propagation is close to rectilinear, and decreases for increasing $R_{\rm s}$ due to the diffusion. The curves rise again at large $R_{\rm s}$, as the particles suffering large deflections have not enough time to reach the observer. 

We provide now some fitting functions for $\langle\cos\theta\rangle$ and $\kappa$ as a function of $R_{\rm s}$ and $d_i$, that are useful to describe the distribution of arrival directions for different physical parameters (distance to the source and emission period, magnetic field amplitude and coherence length, energy and charge of the particles) 
without the need to perform new simulations for each case.  

In the diffusive regime, the dipole amplitude is related to the density through $\Delta=l_D\nabla n/n$, and from this relation a good fit to
 $\langle\cos\theta\rangle$ 
 was obtained in  \cite{mr19} for the case of a steady source. That fit can be slightly modified so that it also applies for finite $d_i$ values, as 
  \begin{equation}
 \langle\cos\theta\rangle\simeq C(R_{\rm s})\left[1+\left(1.6-\frac{3}{2d_i}\right)\left(\frac{R_{\rm s}^2}{0.7 d}\right)^{0.8-0.5/d_i}\right]\equiv C'(R_{\rm s},d_i).
 \end{equation}
 This expression is accurate as long as $R_{\rm s}<d_i/2$, in which case the deflections of the particles are large. One may further improve the agreement with the simulations by requiring that in the limit $R_{\rm s}\to d_i$, where only particles suffering very little deflections can reach the observer, one should have that $\langle\cos\theta\rangle\to 1$.  A reasonable fit to the results of the simulations can be obtained with the expression 
 \begin{equation}
 \langle\cos\theta\rangle \simeq C'(R_{\rm s},d_i)+(1-C'(R_{\rm s},d_i))(R_{\rm s}/d_i)^3,
 \label{eq:cosqg2}
 \end{equation}
as shown in Fig.~\ref{fig:fitvsr}. This expression is valid for all values of $R_{\rm s}$ but as long as $d_i>2$. For smaller values, i.e. when the particles travelled less than a few diffusion lengths, the above expression turns out to overestimate the actual value of $\langle\cos\theta\rangle$. 

For shorter maximum emission times, when $d_i \leq 2$, a good fit is given by
\begin{equation}
 \langle\cos\theta\rangle (R_{\rm s},d_i) \simeq \exp\left(-\frac{R_{\rm s}(1+R_{\rm s})}{3}\right)+\left(1-\exp\left(-\frac{R_{\rm s}(1+R_{\rm s})}{3}\right)\right)\left(\frac{R_{\rm s}}{d_i}\right)^{3.7/d_i}.
 \label{eq:cql2}
 \end{equation}
  
 The concentration parameter $\kappa$ can be fitted by adding to the steady result given in eq.~(\ref{eq:kappasteady}) a term describing the observed growth  as $R_{\rm s}$ approaches $d_i$, with the expression
 \begin{equation}
 \kappa(R_{\rm s},d_i)\simeq \kappa^{\rm steady}(R_{\rm s})
 +\frac{0.44}{(d_i/R_{\rm s})^{0.8+0.4/d_i}-1}.
  \label{eq:fitkappa}
 \end{equation}
 In Fig.~\ref{fig:di.dist} we show, with dashed lines, the Fisher distributions  in eq.~(\ref{eq:fisher}) with values of $\kappa$ and $i$ as given by eqs.~(\ref{eq:fitkappa}) and (\ref{eq:i}), using eq.~(\ref{eq:cosqg2}) or (\ref{eq:cql2}) depending on the value of $d_i$, for the two values of $R_{\rm s}$ reported. It can be seen that the curves provide a good description of the simulated distribution in all cases in the region where the density of particles is significant. Small differences  appear only in some cases when considering backward directions with respect to the source, which are associated to very low fluxes when the propagation is quasi-rectilinear.    
 
\begin{figure}[t]
\centering{
\includegraphics[scale=1.1,angle=0]{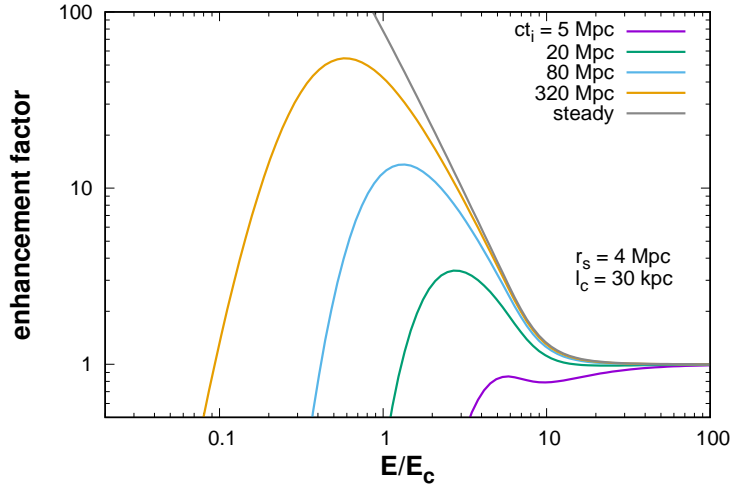} }
\caption{Enhancement factor $\xi$ for a source at a distance $r_{\rm s} = 4$~Mpc, and for different extension of the emission period as a function of $E/E_{\rm c}$. A coherence length of 30~kpc was adopted for the turbulent magnetic field. }
\label{fig:xivse}
\end{figure}

 From the values of the enhancement factor $\xi$,  $\langle\cos\theta\rangle$ and $\kappa$ as a function of $R_{\rm s}$ and $d_i$, we can describe the expected density and angular distribution of arrival directions for any situation by specifying the magnetic field, source distance and emission time of interest. As an example, we show in Fig.~\ref{fig:xivse} the enhancement factor that allows to obtain the spectrum, for a source at a distance of 4~Mpc, emitting since different initial times (as labelled in the plot) in the presence of a turbulent magnetic field with  a coherence length equal to 30~kpc.  The enhancement factor is plotted as a function of $E/E_{\rm c}$, which means that the spectrum of particles with the same rigidity experience the same enhancement. The enhancement factor tends to unity at high energies, where the propagation is quasi-rectilinear, and it increases for lower rigidities due to the diffusion, and finally drops  at the lowest rigidities due to the magnetic horizon effect. The maximum of the enhancement is attained at the energy for which $l_D(E_{\rm max}/E_{\rm c})\simeq 1.1~r_{\rm s}^2/ct_i$, and the enhancement factor at that energy is $\xi_i^{\rm max}\simeq 0.8ct_i/r_{\rm s}$ \cite{mr19}.
 The peak will thus appear at higher energies for heavier nuclei.  If the source were to emit a mixed composition, with the same spectral shape for each component as a function of the rigidity, the enhancement factor found would then lead to an increase in the average mass number as the energy increases.
 
\begin{figure}[t]
\centering{
\includegraphics[scale=.85,angle=0]{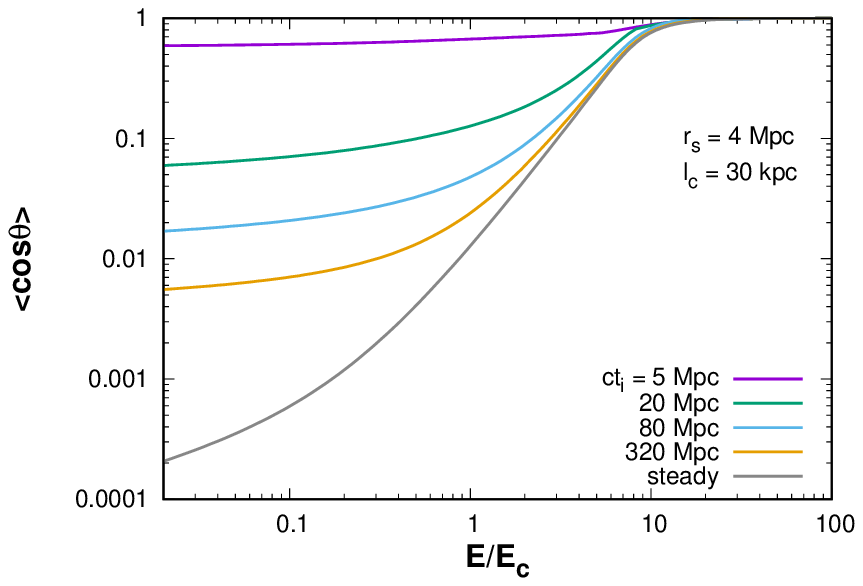}\includegraphics[scale=.85,angle=0]{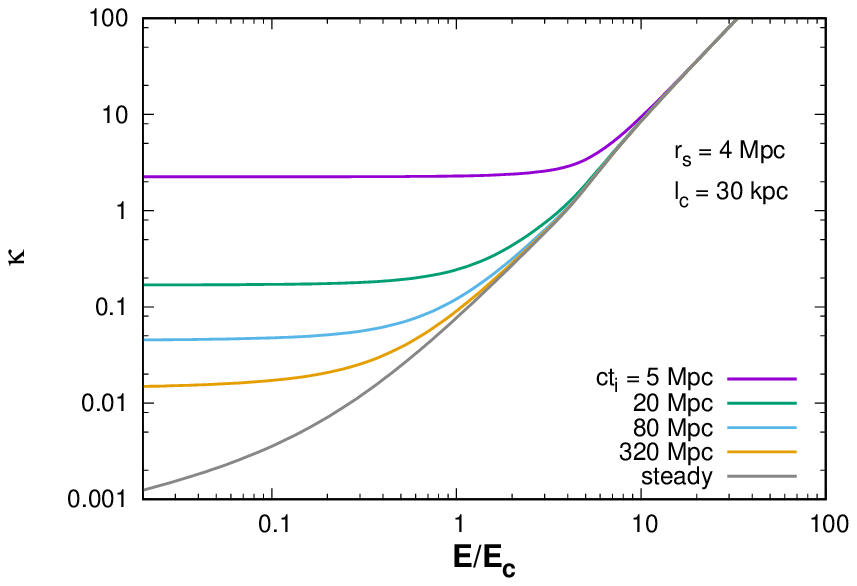} }
\caption{Values of $\langle\cos\theta\rangle$ and $\kappa$ as a function of $E/E_{\rm c}$ for the same values of the source distance and coherence length adopted in Fig.~\ref{fig:xivse}. }
\label{fig:cqkapvse}
\end{figure}

In Fig.~\ref{fig:cqkapvse} we show the values of $\langle\cos\theta\rangle$ and $\kappa$  as a function of $E/E_{\rm c}$ for the same source distance and coherence length considered in Fig.~\ref{fig:xivse}. These parameters are relevant to obtain the expected dipolar amplitudes and the anisotropies on smaller angular scales, respectively. 

Notice that the total dipole will be a superposition of the contributions from all the individual sources, where each nuclear component $j$ of a source in the direction $\hat k_i$ contributes with $\vec\Delta_i^{(j)}(E)= 3 \langle \cos\theta \rangle_i^{(j)} \hat k_i$, where $\langle \cos\theta \rangle_i^{(j)}$ is the mean cosine angle around the source position for nuclei $j$ with energy $E$. Then, if $f_j$ is the fraction of the source flux emitted as nuclei of type $j$ (in a differential energy bin around $E$), considered to be the same for all sources for simplicity, the total dipole can be obtained as
\begin{equation}
\vec\Delta(E)=\sum_{i,j} f_j \frac{n_i^{(j)}(E)}{n_t(E)} \vec\Delta_i^{(j)}(E),
\end{equation}
 where  $n_t(E)=\sum_{i,j}f_j n_i^{(j)}(E)$, with $n_i^{(j)}(E)$ being the density at the observer's position that would result if the source $i$ were just emitting nuclei of type $j$.  From the left panel of Fig.~\ref{fig:cqkapvse} we see that if there is a significant contribution of cosmic rays coming from a very local source, the time since it started to emit cannot be too short in order that the dipole anisotropy does not exceed the values of  few $\%$ ($<10$\%) observed in the energy range from 4 to 30~EeV \cite{lsra19}. 
 
 Regarding the anisotropies at smaller angular scales, since 
$\langle{\theta^2}\rangle\simeq 2/\kappa$, only for values $\kappa > 2$ they are expected to be present. For the example shown in the right panel of Fig.~\ref{fig:cqkapvse}, this can only be expected to happen for energies larger than about $ 5 E_{\rm c}$.

\section{Discussion}
We have considered CR propagation  through the turbulent extragalactic magnetic fields, and studied the effects on the spectrum and anisotropies that result when the source emission is transient. This can result either from a burst in the source activity or be due to the finite time elapsed since a continuous emission started. We considered the main changes that take place in the diffusive regime, and then focused in the transition to the quasi-rectilinear regime, which is the situation in which one expects to see more localized CR flux excesses around the source direction. We also compared the results obtained with the ones usually considered, that correspond to steady sources emitting for very long times. The main results of this study are:

\begin{itemize}
    \item The finite time of the emission leads in general to a suppression of the spectrum at low energies, an effect that is usually referred to as the `magnetic horizon' suppression. This is because at low energies it can take a time much longer than the age of the source for the diffusing particles to reach the observer, and hence essentially no CR flux  is observed from the source. This feature can be helpful to account for the apparently very hard spectrum associated with each observed mass component at ultrahigh energies.
     
        \item In the case of a bursting source, also a suppression appears at the high-energy end due to propagation effects, leading to a spectral density peaked when $l_D(E/E_{\rm c})\simeq r_{\rm s}^2/2ct$ (see Fig.~\ref{fne}). The fact that  the peak appears at higher energies for higher mass components can give rise to scenarios that could explain the spectrum and composition observations at the highest energies  \cite{mr19}. A detailed comparison with the experimental results, including the expected anisotropies, would constrain the relevant parameters characterizing the bursting source and magnetic field.

\item For the bursting source,  only if the travelled distance $ct$ is slightly larger than $r_{\rm s}$ one can expect that the arrival directions would be strongly concentrated around the source direction, having a typical spread $\bar\theta\simeq \sqrt{1-\langle \cos\theta\rangle}\simeq 1.1(d/R_{\rm s}-1)$, so that for instance for $d=1.3R_{\rm s}$  one has that $\bar\theta\simeq 20^\circ$.
    
    \item If the propagation time is larger, one may actually have a deficit in the CR flux around the source direction at the energies for which the diffusion length becomes comparable or larger than the distance to the source ($R_{\rm s}<1$).  In this situation, the CRs would actually arrive preferentially sideways with respect to the direction to the source, and in some cases the CRs may even arrive preferentially from the opposite hemisphere with respect to the source, as is apparent from the negative values of $\langle\cos\theta\rangle$ appearing in Fig.~\ref{fig:burst.cos.16}  for $d\leq 2$. This is because in these conditions the CRs typically make less than a whole turn in the available time,  but those travelling straighter from the source  have already passed through the Earth in the past.
    
    \item When the burst time is farther in the past, such that the distance travelled is much larger than both the diffusion length ($d\gg 1$) and  the source distance ($d\gg R_{\rm s}$), the observed CR distribution acquires an approximately dipolar shape, with $\Delta\simeq 1.5R_{\rm s}/d$, and  the quadrupolar component is subdominant, with $q/\Delta\simeq 7.5(R_{\rm s}/d)^3$.
    
    \item For a source emitting continuously since a given initial time $t_i$, one has that the distribution is always peaked towards the source direction.  It can generally be described with a Fisher distribution, except possibly for backward directions when the propagation from the source is quasi-rectilinear, in which case almost no particles  can arrive from directions opposite to that of the source.
    
    \item It is useful to view this case as a succession of many bursts, since the initial emission time up to the present, and the contribution which is more localized  towards the source direction would be that emitted later, involving travel times only slightly larger that of straight ropagation from the source, while those emitted earlier should arrive more isotropically distributed (as long as $d_i-R_{\rm s}\gg 1$).
    
    \item It is clear that if the emission were not constant in time after the source started its activity, the relative weight of the different `bursting episodes' in the above picture would be affected and hence the final appearance of the CR distribution would be accordingly modified.  For instance, an increased emission in more recent times would make the  source appear more point-like.
    
    \item If a localised excess in the CR arrival distribution were to be detected with large significance at the highest observed energies, the results obtained in this paper could  be useful to better characterize  the source emission history. Given that there are indications that the CR fluxes may consist of a superposition of different nuclear charges, the picture would be further complicated by the combination of the different images of each nuclear component. In some cases, such as in that of a bursting source, the almost independence of the anisotropy signal with energy in the diffusive regime could however simplify the analysis. In this case, the superposition of the dipolar pattern from the source, which is similar for all components as long as they diffuse, with an isotropic background population, could result in a dipolar pattern that will just change with energy due to the energy dependence of the relative contribution of the bursting source to the overall CR flux. Moreover, a departure from a dipolar pattern would be expected  for $d<2\pi$, i.e. if the propagation is close to the quasi-rectilinear regime, in which case the CR distribution may turn out to actually be enhanced along directions away from the source location.

\end{itemize}

\section*{Acknowledgments}
This work was supported by CONICET (PIP 2015-0369) and ANPCyT (PICT 2016-0660).


\end{document}